\documentstyle[12pt,epsf]{article}
\newcommand{\bfig}{\begin{figure}}
\newcommand{\efig}{\end{figure}}

\def\ap{\alpha^{\prime}}
\def\p{\partial}

\def\zb{\bar{z}}
\def\zp{z^{\prime}}
\def\zpb{\bar{z}^{\prime}}
\def\Psib{\bar{\Psi}}
\def\Phib{\bar{\Phi}}
\def\db{\bar{d}}
\def\bb{\bar{b}}
\def\12{\frac{1}{2}}
\def\bea{\begin{eqnarray}}
\def\eea{\end{eqnarray}}
\def\ba{\begin{array}}
\def\ea{\end{array}}
\def\vt{\vartheta}

\def\one-loop{\mbox{\scriptsize one-loop}}

\def\a{\alpha}
\def\ab{\bar{\alpha}}
\def\b{\beta}

\def\s{\sigma}
\def\th{\theta}

\def\psib{\bar{\psi}}

\def\G{\Gamma}

\catcode`\@=11

\@addtoreset{equation}{section}
\def\theequation{\arabic{section}.\arabic{equation}}

\def\@normalsize{\@setsize\normalsize{15pt}\xiipt\@xiipt
\abovedisplayskip 14pt plus3pt minus3pt%
\belowdisplayskip \abovedisplayskip
\abovedisplayshortskip  \z@ plus3pt%
\belowdisplayshortskip  7pt plus3.5pt minus0pt}

\def\small{\@setsize\small{13.6pt}\xipt\@xipt
\abovedisplayskip 13pt plus3pt minus3pt%
\belowdisplayskip \abovedisplayskip
\abovedisplayshortskip  \z@ plus3pt%
\belowdisplayshortskip  7pt plus3.5pt minus0pt
\def\@listi{\parsep 4.5pt plus 2pt minus 1pt
            \itemsep \parsep
            \topsep 9pt plus 3pt minus 3pt}}

\def\underline#1{\relax\ifmmode\@@underline#1\else
        $\@@underline{\hbox{#1}}$\relax\fi}
\@twosidetrue

\relax

\catcode`@=12

\catcode`\@=11

\def\section{\@startsection{section}{1}{\z@}{3.5ex plus 1ex minus
   .2ex}{2.3ex plus .2ex}{\large\bf}}

\def\thesection{\Roman{section}.}

\def\appendix{\setcounter{section}{0}
        \def\thesection{Appendix }
        \def\theequation{\Alph{section}.\arabic{equation}}}

\def\ps@headings{\def\@oddfoot{}\def\@evenfoot{}
\def\@oddhead{\hbox{}\hfill
        \makebox[.5\textwidth]{\raggedright\ignorespaces --\thepage{}--
        \hfill {}}}
\def\@oddhead{\hbox{}\hfill --\thepage{}-- \hfill
        {}}
\def\@evenhead{\@oddhead}
\def\subsectionmark##1{\markboth{##1}{}}
}

\ps@headings

\catcode`\@=12

\relax

\def\figcap{\section*{Figure Captions\markboth
        {FIGURECAPTIONS}{FIGURECAPTIONS}}\list
        {Fig. \arabic{enumi}:\hfill}{\settowidth\labelwidth{Fig. 999:}
        \leftmargin\labelwidth
        \advance\leftmargin\labelsep\usecounter{enumi}}}
 \relax
\def\tablecap{\section*{Table Captions\markboth
        {TABLECAPTIONS}{TABLECAPTIONS}}\list
        {Table \arabic{enumi}:\hfill}{\settowidth\labelwidth{Table 999:}
        \leftmargin\labelwidth
        \advance\leftmargin\labelsep\usecounter{enumi}}}
 \relax
\def\reflist{\section*{References\markboth
        {REFLIST}{REFLIST}}\list
        {[\arabic{enumi}]\hfill}{\settowidth\labelwidth{[999]}
        \leftmargin\labelwidth
        \advance\leftmargin\labelsep\usecounter{enumi}}}
 \relax

\catcode`\@=11

\def\ps@headings{\def\@oddfoot{}\def\@evenfoot{}
\def\@oddhead{\hbox{}\hfill
        \makebox[.5\textwidth]{\raggedright\ignorespaces --\thepage{}--
        \hfill {}}}
\def\@evenhead{\@oddhead}
\def\subsectionmark##1{\markboth{##1}{}}
}

\ps@headings

\relax

\newskip\humongous \humongous=0pt plus 1000pt minus 1000pt

\newif\ifdtup

\def\beq{\begin{equation}}
\def\eeq{\end{equation}}

\def\beqn{\begin{eqnarray}}
\def\eeqn{\end{eqnarray}}
\relax

\def\G2{{\; \rm GeV/}c^2}
\def\G{\; \rm GeV}

\def\dotx{\dotx{\dot\overline{x}}}

\relax

\def\p{\partial}

\begin{document}
\begin{titlepage}
\begin{flushright}
       {\normalsize OU-HET 330 \\  hep-th/9910263 \\}
\end{flushright}

%
\begin{center}
  {\large \bf $p-p^\prime$ System with $B$ Field, \\
   Branes at Angles and  Noncommutative Geometry}
\footnote{This work is supported in part
 by the Grant-in-Aid  for Scientific Research (10640268)
 and Grant-in-Aid  for Scientific Research fund (97319)
from the Ministry of Education, Science and Culture, Japan,
and in part by
the Japan Society for the Promotion of Science for Young
Scientists.}

\vfill
         {\bf  B. Chen, H. Itoyama, T. Matsuo and K. Murakami\footnote{Email address: chenb, itoyama, matsuo, murakami@funpth.phys.sci.osaka-u.ac.jp.}}\\
\vfill
        Department of Physics,\\
        Graduate School of Science, Osaka University,\\
        Toyonaka, Osaka 560-0043, Japan\\
\end{center}
\vfill
\begin{abstract}

We study the generic $p-p^\prime$ system in the presence of
constant NS 2-form $B_{ij}$ field. 
We derive properties concerning with the noncommutativity of
D-brane worldvolume, the Green functions and the spectrum of this system.
In the zero slope limit, a large number of light states
appear as the lowest excitations in appropriate cases.
We are able to relate the energies of the lowest states
after the GSO projection with the configurations of
branes at angles.
Through analytic continuation, the system is compared with the branes
with relative motion.
\end{abstract}
\vfill
\end{titlepage}

\section{Introduction}
  There are several reasons to expect that
noncommutative geometry
may play a pivotal role in the developments of string theory. 
The developments  which have already taken place include  an effective
worldvolume description of D-branes in terms of noncommuting
coordinates \cite{Witten2}. One may mention
the recent proposal by Connes, Douglas and Schwarz that, in the small
volume limit, the 
compactification of Matrix theory on a torus with background $C_{-ij}$
should be described by noncommutative Yang-Mills theory on the dual 
torus \cite{CDS, DH}. The T-duality in string theory is found
to have a counterpart in noncommutative geometry called Morita
equivalence \cite{AS}.
Much work has been done, following these developments. (See \cite{SW} for
 references.) Let us mention here  the work on
 instantons on noncommutative space. In order to remove the
 singularities of the instanton moduli
space, one needs to introduce a nonvanishing parameter in the ADHM
construction of instantons \cite{Naka}. Nekrasov and Schwarz noticed
that the ADHM construction of instantons on noncommutative space recovers
 this modification so that the smooth moduli of the instantons are the moduli
 space of instanton on noncommutative space\cite{Nek}. 
 
These ideas have recently been extended by Seiberg and Witten \cite{SW}. 
 Noncommutativity of D-brane worldvolumes has been shown to emerge from
the open string Green function in the presence of  constant rank-two
  antisymmetric background field $B_{ij}$. In fact,
at the boundaries of an open string, the two-point Green function is 
\beq
\left\langle x^i(\tau)x^j(\tau^\prime) \right\rangle
  =-\ap G^{ij}\log(\tau-\tau^\prime)^2+
   \frac{i}{2}\theta^{ij}\epsilon(\tau-\tau^\prime) \;\;,
\eeq
where $\tau \equiv \mid z \mid $, $G^{ij}$ is an open string metric
 and $\theta^{ij}$ is a noncommutative parameter measured by
\beq
[x^i(\tau), x^j(\tau)]=i\theta^{ij}\;\;\;.
\eeq
As mentioned above, the D-brane worldvolume is a noncommutative space. 
In the zero slope limit $\ap \to 0$ while keeping the open string
 metric $G^{ij}$ and the noncommutative parameter $\theta^{ij}$ finite,
 the operator product becomes a $\ast$ product of noncommutative geometry,
 which is associative and translationally invariant.
 Properties of the D-branes can be described by an effective action on  the
 noncommutative space. Several other topics have been investigated in \cite{SW}. 

In the D0-D4 system with $B_{ij}$ background, in particular, it has been found that 
 a large number of light states emerge when $Pf(B)<0$.
In the zero-slope limit,
the system can be effectively considered as the configuration in which there
 is a continuous distribution of D0's in the D4 worldvolume. The properties
 of the system can then be treated as those of D0-D0 pairs. The existence of
 a large number of light states could be interpreted as fluctuations of
 the one end point of the open string stretching between two D0s, keeping the
 other end fixed. As the D0-D4 system without $B_{ij}$ field can be
 understood in terms of small instantons \cite{Witten1, Douglas}, the low lying 
D0-D4 excitation spectrum with $B_{ij}$ should correspond to small fluctuations
 around point-like instantons. But now the instantons live in noncommutative $R^4$. 

In this paper, we consider the generic $p$-$p^\prime$ system in
the presence of  $B$ field.  In the next section, we will derive the value of
 the parameter $\theta$
which measures noncommutativity, the Green function
 and the excitation spectrum. In \cite{SW}, it has been shown that
 in the presence of constant $B_{ij}$
 the D-brane worldvolume should 
be noncommutative. The analysis is based on the open string Green function in the background \cite{Callan1}. The 
end point of the open string probes the worldvolume structure of D-brane.
 This suggests that we may investigate
the D$p$ worldvolume via D$p$-D$p^\prime$ open string.
We find that although the open string Green function is 
quite different from the one in eq.(1.1) where only D$p$-branes
are considered, the noncommutative 
structure is the same. Our calculation gives a consistency check of the
formalism: in D$p$-D$p^\prime$ system, 
we can study the space structure either with D$p$-D$p$ open string or with 
D$p$-D$p^\prime$ open string. The two approach 
should give the same answer. This might indicate the existence of
 a universal description of the 
noncommutativity of D$p$-brane worldvolumes rather than the one
 by the open string.        
We find that
a number of light states emerge in the zero slope limit
in some appropriate cases. 
Since the D0-D2 system in the presence of the $B$ field  can be used as
 a building block of the Green functions and the description of the states
 in terms of the modes,  we will discuss  this case in some detail.

Another theme on the $p$-$p^\prime$ system
with  constant $B$ field which we
observe is its equivalence to the system of two  D-branes at
angles after T-duality. 
Configurations of D-branes at angles have been studied
in \cite{Berkooz, Leigh}.
It has been noticed that in general such configurations  break all of  the
supersymmetries and only for some special configurations
a fraction of them survives. One such  example is
two D2 branes intersecting at right angles. Via T-duality, it
is related to the D0-D4 system. We provide appropriate
identifications between $\nu$ s in the $p$-$p^\prime$ system  originated
from the canonical form of $B$  and angles in the T-dualized system,
paying attention to the GSO projection and BPS configuration.
It is found that the energies of the ground state  after the GSO projection
 and the lowest
excitations in the $p$-$p^\prime$ system are closely related
to the condition for supersymmetry in the system of branes at angles.
 The above identifications measure how far the system is from a BPS
 configuration.
In the dual picture, the light states correspond to 
the small fluctuations around the BPS configuration.
For example, in the case of D0-D4,  under our identification,
the energies of the lowest GSO states are proportional to
$\phi_1+\phi_2$.
When $Pf(B)<0$, the T-dualized picture is a D2 almost parallel to
another D2, in the $\ap\to 0$ limit, and
a large number of light states appear. And when $Pf(B)>0$,
the T-dualized picture
is two almost anti-parallel D2 branes. The tachyon appears as the lowest
state and has energy proportional to $\pi$. 
There are no light states in the limit of $\ap \to 0$.
The value of $\phi_1+\phi_2$   tells us whether we are near or 
far from a BPS configuration. 
In fact, even without taking the zero slope limit, we can 
know from  the dual picture in which case  our $p$-$p^\prime$ system 
 keeps a fraction of supersymmetries.
In the case of D0-D4, as long as $B_{12}=-B_{34}$, 
the identification leads to $\phi_1+\phi_2=0$,
which is the condition for supersymmetry in the case of 
two D2 branes at angles, so that our system is BPS 
although the worldvolume of D4 now is noncommutative.
In the case $\overline{\rm D0}$-D4 with $B_{ij}$,
the BPS condition requires that  $B_{12}=B_{34}$.
The careful treatment of the generic D0-D$p$ $(p=2,4,6,8)$ system
will be discussed in section III.

 In section IV, we compare  our $p$-$p^\prime$ system with $B_{ij}$
background with two D-branes with relative motion. 
The dynamics of D-branes with relative motion has also been noted for some
time \cite{Bachas1, Lif}. 
It has been related to an open string pair production in a
 constant electric background \cite{Bachas2}. 
Therefore, for generic $B_{\mu\nu}$ background, the $p$-$p^\prime$
system can be 
treated  either as the two D-branes with relative motion or
as those with relative orientation  depending upon whether the time direction
is included or not.

The one-loop vacuum amplitude has been evaluated. 
  The dependence on  $\nu$ tells
us in which case
we have a  nearly supersymmetric configuration. 
The effect of $B_{\mu\nu}$ can be seen from the investigation 
of either motion D-branes or D-branes at angles.
This means the dynamics of D-branes come 
from the introduction of $B_{\mu\nu}$ field.


\section{$p$-$p^\prime$ System with $B_{ij}$ Field}

In \cite{SW}, it has been shown that two D$p$-branes in the presence of
constant $B_{ij}$ field can still preserve one-half of the
supersymmetries and the excitations from an open string between
the two D$p$-branes are unchanged.
The only effect of $B$ field is to make its worldvolume noncommutative.
In order to describe the D-brane worldvolume geometry,
it is expedient to introduce the parameter $\theta_{ij}$  which measures
 noncommutativity
and the open string metric $G_{ij}$.
In the zero-slope limit, the tower of string excitations are projected
out and we obtain an effective description in terms
of finite $\theta$ and $G$. 
The low energy effective action of the D$p$-brane
is super-Yang-Mills theory defined on a noncommutative space \cite{SW}.

In the case of $p$-$p^\prime$ system,
we find that supersymmetry can not be kept generically
and tachyons are ubiquitous.
As  $B_{ij}$  can be set  to its canonical form
\beq
B_{ij}=\frac{\epsilon}{2\pi\ap} \left(\ba{ccccc}
0&b_{1}&0&0&0\\
-b_{1}&0&0&0&0\\
0&0&0&b_{2}&\vdots\\
0&0&-b_{2}&0&\vdots\\
0&0&\cdots&\cdots&0
\ea\right) \;\;,
\label{eq:bfield}
\eeq
and the space-time is flat with metric
\begin{equation}
 g_{ij} = \epsilon \delta_{ij}~,\label{eq:metric}
\end{equation}
we can reduce the problems of the mode expansions and the Green functions
in the $p$-$p^\prime$ system to those in the D0-D2 system.
Let us focus on this case for a while.

We take the D2 worldvolume to extend in the
 $x^0, x^1, x^2$ directions. We define $Z \equiv x^1+ix^2$.
The boundary conditions on the D0-D2 bosonic open strings
$Z$ and $\overline{Z}$ are 
\bea
\p_{\tau} Z |_{\s=0}
  &=&\p_\s Z+b\p_{\tau} Z |_{\s=\pi}=0 \;\;, \nonumber \\
\p_\tau \overline{Z}|_{\s=0}
 &=&\p_\s \overline{Z}-b\p_\tau \overline{Z}|_{\s=\pi}=0
 \;\;.
\label{eq:d2bc}
\eea
The mode expansion of $Z$ and that of $\overline{Z}$ satisfying
the above boundary conditions are
\bea
\label{eq:mode}
Z&=&i\sqrt{\frac{\alpha^{\prime}}{2}}
    \sum_n(z^{n+\nu}-\zb^{n+\nu})\frac{\a_{n+\nu}}{n+\nu} \;\;, \\
\overline{Z}&=&i\sqrt{\frac{\alpha^{\prime}}{2}}
    \sum_n(z^{n-\nu}-\zb^{n-\nu})\frac{\ab_{-n+\nu}}{n-\nu} \;\;,
\nonumber
\eea
where
\beq
e^{2\pi i\nu}=-\frac{1+ib}{1-ib}\;,\hspace{5ex} 0\leq \nu <1 \;\;,
\eeq
and $z=e^{\tau+i\sigma}$ $({\rm Im}z \geq 0)$.
When $b$ is small or goes to $\pm \infty$, $\nu$ can be approximated by
\beq 
\nu \approx \left\{ \ba{ll}
   {}-\frac{1}{\pi b},&b\to -\infty\\
     \frac{1}{2}+\frac{b}{\pi},&b\approx 0\\
     1-\frac{1}{\pi b},&b\to \infty
\ea \right. 
\eeq

{}From quantization, we know the commutation relation of $\a$ and $\ab$:
\beq
\label{eq:crel}
[\a_{n+\nu},\ab_{-m+\nu}]=-\,\frac{2}{\epsilon}\,(n+\nu)\delta_{m+n} \;\;.
\eeq
  The oscillator ground state $|0\rangle $ implements the conditions
\beq
\a_{n+\nu}|0\rangle=0  \;\;, \hspace{5ex} n<0 \;\;,
\eeq
\beq
\ab_{m+\nu}|0\rangle=0\;\; , \hspace{5ex} m\geq 0 \:\:.
\eeq
in the current notation. (Notice the minus sign of the
 right hand of eq. (\ref{eq:crel}) ).
 We have two groups of creation operators, the one consisting of $\a_{n+\nu}$
 with $n\geq 0$ and  the other of $\ab_{m+\nu}$ with $m<0$. 
We will find that the excitation energies are different
between these two groups.

The Virasoro generators of the $Z$-$\overline{Z}$ system are given by 
\beq
L_m = \frac{\epsilon}{2}\sum_n :\a_{n+\nu}\ab_{m+n+\nu}: \;\;.
\eeq
The ground state energy is
\beq
E(\nu)=\sum_{n=1}^\infty (n-\nu)
 =\frac{1}{24}-\frac{1}{2}\left(\nu-\frac{1}{2}\right)^2\;\;.
\eeq
{}From the eigenvalues of $L_{0}$ action,
we can read off the excitation energies.
The two lowest excitations are 
$\a_\nu$ with its energy $\nu$ and $\ab_{-1+\nu}$ with its energy $1-\nu$.
  Which one is the lowest depends on whether $\nu$ is bigger or smaller
 than one-half, or equivalently on whether $b$ is positive
 or negative.

%
%

We have the D0-D2 open string with one end on the D0 brane and the other
on the D2 brane. To evaluate the parameter $\theta$  which measures the
noncommutativity on the D2 worldvolume, let us 
find the commutator between $Z$ and $\overline{Z}$ at the D2 endpoint. 
This leads us to calculate the two-point
function between $Z$ and $\overline{Z}$. From the mode expansion
eq.\ (\ref{eq:mode}), we find
\bea
\langle 0| Z(z)\overline{Z}(\zp) |0\rangle
 &=& - \frac{\alpha^{\prime}}{2}
       \langle 0 |\sum_{n<0}\frac{\a_{n+\nu}}{n+\nu}
        \left( z^{n+\nu}-\zb^{n+\nu}  \right) 
      \sum_{m>0}\frac{\ab_{-m+\nu}}{-m+\nu}
        \left(z^{\prime m-\nu}-\bar{z}^{\prime m-\nu}\right)
     |0 \rangle \nonumber\\
&=&-\frac{\alpha^{\prime}}{\epsilon}\sum_{n<0}\frac{1}{n+\nu}\left[
     \left(\frac{z}{\zp}\right)^{n+\nu}
    {}-\left(\frac{z}{\zpb}\right)^{n+\nu}
    {}-\left(\frac{\zb}{\zp}\right)^{n+\nu}
     +\left(\frac{\zb}{\zpb}\right)^{n+\nu}\right] \;\;,
\eea
where  the infinite series can be written in terms of the
  hypergeometric function
\beq
\Phi(z,1,\nu)=\sum_{n=0}^{\infty}\frac{z^n}{n+\nu}=
\nu^{-1} {_2F_1}(1,\nu;1+\nu;z)\;, \hspace{5ex}|z|<1 \;\;.
\eeq
Similarly, we can obtain the two-point function
$\langle 0|\overline{Z}(z)Z(\zp)|0 \rangle $.
The commutator at the end point  $\s=\pi$  is
\bea
{[Z,\overline{Z}]}& = &-\frac{\ap}{\epsilon}\sum_{n\in Z}
\frac{1}{n+\nu}(2-e^{2i\nu\pi}-e^{-2i\nu\pi})=\frac{2\pi\ap}
{\epsilon}\cdot\frac{2b}{1+b^2}~,  \nonumber \\
{[x^1, x^2]}& = &i\frac{2\pi\ap}{\epsilon}\cdot \frac{b}{1+b^2}~.
\label{eq:com}
\eea
Eq.\ (\ref{eq:com}) can be compared with  eq.\ (2.5) of \cite{SW}:
\beq
\th^{ij}=2\pi\ap\left(\frac{1}{g+2\pi\ap B}\right)^{ij}_A \;\;,
\eeq
where $(~~)_{A}$ denotes the antisymmetric part of the matrix.
{}Substituting eqs.\ (\ref{eq:bfield}) and (\ref{eq:metric}), one has
\beq
\label{eq:comsw}
\th^{12}=\frac{2\pi\ap}{\epsilon}\cdot \frac{b}{1+b^2}~.
\eeq
Eq.\ (\ref{eq:com}) and eq.\ (\ref{eq:comsw}) agree.    The Green
 function of the D$p$-D$p$ system  and that of the D0-D2 system have
 the same noncommutativity.
 
The open string metric $G_{ij}$ on the D2 worldvolume can be
extracted from the Green function evaluated at the end point $\s=\pi$.
We find 
\beq
\label{eq;zz}
\left.\langle 0|Z(z)\overline{Z}(\zp)|0\rangle \right|_{\s=\pi}
=\frac{\alpha^{\prime}}{\epsilon}\sum_{n=1}^{\infty}\frac{1}{n-\nu}
\left(\frac{e^{\tau^{\prime}}}{e^{\tau}}\right)^{n-\nu}
\left(2-e^{2i\nu\pi}-e^{-2i\nu\pi}\right) \;\;.
\eeq
From the property 
\beq
\lim_{z\to 1}\Phi(z,1, \nu)/[-\log(1-z)]=1 \;\;,
\eeq
  we find that the singular behaviour of eq. (\ref{eq;zz})
 as $\tau \to \tau^\prime$ is logarithmic:
\begin{eqnarray}
&&\left. 
  \langle 0 | Z(z) \overline{Z}(z^{\prime}) |0\rangle 
\right|_{\sigma=\pi}
\sim -\frac{\alpha^{\prime}}{2\epsilon}
     \left( 2-e^{2i\nu\pi}-e^{-2i\nu\pi}\right)
     \log \left( e^{\tau}-e^{\tau^{\prime}}\right)^{2}~,\nonumber\\
&&\left. 
     \langle 0 | x^{1}(z)x^{1}(z^{\prime}) | 0 \rangle
   \right|_{\sigma=\pi}
=  \left. 
     \langle 0 | x^{2}(z)x^{2}(z^{\prime}) | 0 \rangle
   \right|_{\sigma=\pi}
   \sim -\frac{\alpha^{\prime}}{4\epsilon}
     \left( 2-e^{2i\nu\pi}-e^{-2i\nu\pi}\right)
     \log \left( e^{\tau}-e^{\tau^{\prime}}\right)^{2}\nonumber\\
&& \hspace{9.4em}
   =-\alpha^{\prime}\frac{1}{\epsilon(1+b^{2})}
     \log \left( e^{\tau}-e^{\tau^{\prime}}\right)^{2}~.
\label{eq:propagate}
\end{eqnarray}
Eq.\ (\ref{eq:propagate}) can be compared with eq.\ (2.5)
in \cite{SW}:
\begin{equation}
 G^{ij}=\left(\frac{1}{g+2\pi\alpha^{\prime}B}\right)^{ij}_{S}~,
\end{equation}
where $(~~)_{S}$ denotes the symmetric part of the matrix.
{}From eqs.\ (\ref{eq:bfield}) and (\ref{eq:metric})
we obtain
\begin{equation}
 G^{11}=G^{22}=\frac{1}{\epsilon(1+b^{2})}~.
\end{equation}
Comparing this with the prefactor of the logarithm
in eq.\ (\ref{eq:propagate}),
we find that the open string metric obtained in our
system is exactly the same\footnote{Please note that the time variables
$\tau$ and $\tau^{\prime}$ in \cite{SW} mean respectively
$e^{\tau}$ and $e^{\tau^{\prime}}$ in this section.}
as that in \cite{SW}.
 
 We conclude that the noncommutativity of the worldvolumes of
 the D-branes and the open string metric can both be probed,
  using the generic   $p$-$p^\prime$ system.

Let us now turn to the worldsheet fermions in the NSR formalism.
 First, define
\beq
\Psi=\psi^1+i\psi^2 \;, \hspace{5ex} \Psib=\psi^1-i\psi^2,
\eeq
\beq
\Phi=\psib^1+i\psib^2,\hspace{5ex} \Phib=\psib^1-i\psib^2 \;\;.
\eeq
The boundary conditions for the worldsheet fermions can be determined
  by demanding the worldsheet supersymmetry.  We find that, in
 the Ramond sector, the worldsheet fermions obey
\beq
\Psi+\Phi|_{\s=0}=(\Psi-\Phi)-ib(\Psi+\Phi)|_{\s=\pi}=0 \;\;,
\eeq
\beq
\Psib+\Phib|_{\s=0}=(\Psib-\Phib)+ib(\Psib+\Phib)|_{\s=\pi}=0 \;\;.
\eeq
The mode expansion on the upper half plane is
\beq
\Psi=\sum_{n\in Z}d_{n+\nu}z^{n+\nu-\frac{1}{2}}\; ,\hspace{5ex}
 \Phi=-\sum_{n\in Z}d_{n+\nu}\zb^{n+\nu-\frac{1}{2}} \;\;,
\eeq
\beq
\Psib=\sum_{n\in Z}\db_{-n+\nu}z^{n-\nu-\frac{1}{2}}\; ,\hspace{5ex}
 \Phib=-\sum_{n\in Z}\db_{-n+\nu}\zb^{n-\nu-\frac{1}{2}}\;\;.
\eeq
The total ground state energy vanishes in the Ramond sector
  due to the cancellation between bosons and fermions.
  As in the bosonic case, the excitations come from two types of
 oscillators $d$ and $\db$, and the two lowest ones have energy 
$E=\nu$ and  $E=1-\nu$ respectively.
 
In the NS sector, the boundary conditions change to 
\beq
\Psi+\Phi|_{\s=0}=(\Psi+\Phi)-ib(\Psi-\Phi)|_{\s=\pi}=0 \;\;,
\eeq
\beq
\Psib+\Phib|_{\s=0}=(\Psib+\Phib)+ib(\Psib-\Phib)|_{\s=\pi}=0\;\;.
\eeq
The mode expansion on the upper half plane reads 
\bea
\Psi &=&   \sum_{r\in Z+\frac{1}{2}}b_{r+\nu}z^{r+\nu-\frac{1}{2}}\;,
\hspace{5ex}
 \Phi=-\sum_{r\in Z+\frac{1}{2}}b_{r+\nu}\zb^{r+\nu-\frac{1}{2}}\;\;,  \\
\Psib &=&  \sum_{r\in Z+\frac{1}{2}}\bb_{-r+\nu}z^{r-\nu-\frac{1}{2}}\;,
\hspace{5ex}
 \Phi=-\sum_{r\in Z+\frac{1}{2}}\bb_{-r+\nu}\zb^{r-\nu-\frac{1}{2}}\;\;.
\eea
  The contribution to the ground state energy from the NS fermions is
 $-E(|\nu-\frac{1}{2}|)$ and in general cannot cancel  the bosonic
 contribution. Two kinds of excitations can be 
ordered  according to the energies they carry:
\beq
\nu > \frac{1}{2}: \; \nu-\frac{1}{2}, \; \frac{3}{2}-\nu,\; \frac{1}{2}+\nu,\;
 \frac{5}{2}-\nu,\; \frac{3}{2}+\nu,\;  \cdots\;.
\eeq
\beq
\nu < \frac{1}{2}: \; \frac{1}{2}-\nu,\; \frac{1}{2}+\nu,\;
 \frac{3}{2}-\nu, \; \frac{3}{2}+\nu,\; \frac{5}{2}-\nu,\; \cdots \;.
\eeq

The total ground state energy in the NS sector is
\bea
E_0&=&3E(0)-3E\left(\frac{1}{2}\right)
      +E(\nu)-E\left(\left|\nu-\frac{1}{2}\right|\right)\nonumber\\
 &=&-\frac{1}{4}-\frac{1}{2}\left| \nu-\12 \right| \nonumber \\
 &=&\left\{\ba{ll}
-\frac{1}{2}\nu&\hspace{3ex}\nu > \frac{1}{2}\\
\frac{1}{2}\nu-\frac{1}{2}&\hspace{3ex}\nu < \frac{1}{2}
\ea \right.  \;\;.
\eea
The first excited state has an energy
\beq
E_1=E_0+\left| \nu-\12 \right|
   =-\frac{1}{4}+\frac{1}{2} \left|\nu-\12 \right| \;\;.
\eeq
 We rewrite the energies of these two states collectively as  
\beq
E^\pm=-\frac{1}{4}\pm\12 \left( \nu-\12 \right)  \;\;,
\eeq
${\it i.e.}$
\beq
\label{eq:pm}
E^+=\12(\nu-1)\;,\hspace{5ex} E^-=-\12\nu \;\;
\eeq
Due to the GSO projection, only one of these two states survive.
We determine which state we should project out in the following way.
As we will disscuss later, D0-branes are induced on the D2-brane
worldvolume in the zero-slope limit \cite{SW}.
When $Pf(B)<0$, we identify them with D0's not $\overline{\rm D0}$'s,
following the argument in \cite{SW}.
Using this definition of the D0-brane, we take the GSO projection
so that the system becomes D0-D2 not $\overline{\rm D0}$-D2.
In the other cases which we will investigate
in the following,
we will take the GSO projection in the same manner. 

By definition, $E^-$ is chosen as the ground state after 
the GSO projection while $E^+$ is left out in the entire region of
 $\nu :\;\;0\leq \nu <1$. We can restate this by saying
 that,  when $b$ is negative, we keep $E_1$ and leave $E_0$ out and
 that, when $b$ is positive, we  keep $E_0$ and leave $E_1$ out.
From this latter point of view, it can be said 
that, in the two regimes of $\nu$, the GSO projections are opposite. 
 We will see that the pattern of classification of states seen
 in eq. (\ref{eq:pm}) will hold in the generic $p$-$p^\prime$ system.

For small positive $b$,  
\beq
E_0\approx -\12\left(\12+\frac{b}{\pi}\right)<0 \;\;,
\eeq
  and we have a tachyonic state  after the GSO projection. 
For small negative $b$, we have 
\beq
E_0=\12\nu-\12\approx-\frac{1}{4}+\frac{b}{2\pi};\;,
\hspace{3ex}E_1\approx -\12 \left( \12+\frac{b}{\pi} \right)\;\;,
\hspace{3ex}E_2
  \approx \frac{3}{2} \left( \12+\frac{b}{\pi} \right)\;\;.
\eeq
The ground state is projected out and the part of the remaining first
excitations is still tachyonic.
 
We are, however, more interested in the situation in the zero slope
 limit, ${\it i.e.}$ $|b|\to \infty$ limit.
We would like to know if some light states survive. This is in parallel to
  the discussion of D0-D4 system in \cite{SW}. 
We will see that a large number of light states appear  for $b$ negative.

As $ b \to + \infty $, the ground state energy becomes
\beq
E_0\approx -\frac{1}{2}\left(1-\frac{1}{\pi b}\right) \;\;,
\eeq
being of order $1$. Clearly this is
  tachyonic  and is the same order  
 as the  energy of the tachyon in the D0-$\overline{\rm D0}$ system. In the
 language of \cite{SW}, the large positive $b$ induces
a large number of $\overline{\rm D0}$s in the D2, and the D0 can annihilate one
 of these $\overline{\rm D0}$s.
All of the one-particle excitations are projected out
and the lowest two-particle excitation has energy
\beq
E=E_0+1=\frac{1}{2} \left(1-\frac{1}{\pi b} \right) \;\;,
\eeq
which is positive, and also of order $1$. 
This state cannot become light in the zero-slope limit.  We conclude that
 our system has no light state  for $b$ positive.

For $ b \to - \infty $, we have a different story. The first two one-particle
 excitations have energies
\beq
E_1\approx \frac{1}{2\pi b}\;,\hspace{2ex}E_2\approx -\frac{3}{2\pi b}\;\;,
\eeq
and their mass squared are finite
as $\ap\to 0$. Apart from these two light states, we obtain
eight more states by acting on $|0 \rangle$ with the creation operators 
which carry the lowest energy $\12$ and 
which are obtained from the NS fermion partners of
 $x^{0,3,\cdots, 9}$.   
These states have energies
\beq
E_i\approx -\frac{1}{2\pi b} \;\;,
\eeq
surviving the $\ap\to 0$ limit. We can further act on these states with an
 arbitrary polynomial consisting of $\a_\nu$  with its energy
 $\nu=-\frac{1}{\pi b}$.  This  also leads to
 a finite energy state in the zero-slope limit.  We conclude that, for $b$
 negative, we have a large number of light states in  the $\ap\to 0$ limit.
   The one end of the open string is fixed and the other end can be 
located anywhere in the D2 worldvolume.  We have a continuous distribution 
of D0s in D2.
 
The appearance of a large number of light states in the zero-slope limit
for $b$ negative can also be understood
in the framework of two D-branes intersecting at angles. 
As we will show in the next section, when $\nu$ is 
about zero, the corresponding angle is about zero. We have
a cofiguration of  two D1 branes with very small relative angle, 
which is nearly supersymmetric.
The excitations are almost massless.

  This completes our discussion of the D0-D2 system with $B_{ij}$ background. 
The D0-D4 system with $B_{ij}$ background has been discussed in \cite{SW} and
 we  will recapitulate this case very briefly for the sake of completeness.
 The four lowest energy states including the ground state
have energies
\beq
E^+_\pm=\pm\12(\nu_1+\nu_2-1)\;,
\hspace{5ex}E^-_\pm=\pm\12(\nu_1-\nu_2)\;\;.
\label{eq:d2energy}
\eeq
  By definition, the GSO projection keeps the states with energies $E^+_\pm$ 
 and  leaves out the states with energies $E^-_\pm$  in the entire region of
  $\nu_1$ and $\nu_2$. When $Pf(B)>0$, the D0-D4 system is tachyonic. 
The tachyon mass squared is of order $1/\ap$ in the $\ap\to 0$ limit.
It implies that this is a standard D0-$\overline{\rm D0}$ tachyon.
When $Pf(B)<0$, and after the
GSO projection,  the lowest state and the first three  excitations survive
 the $\ap\to 0$ limit.   As in the D0-D2 case, the one-particle excitations
 from the lowest fermionic partners of the other directions and the
 multiparticle excitations of arbitrary varieties consisting of
  the lowest bosonic mode
 give light states in the zero-slope limit. We have a large number of
 light states here.

  Let us now turn to the D0-D6 case.
In the  D0-D6 system with $B_{ij}$ field, the ground state in the NS sector has energy 
\beq
E_0=\frac{1}{4}-\12\left(\left|\nu_1-\12\right|
           +\left|\nu_2-\12\right|+\left|\nu_3-\12\right|\right) \;\;.
\eeq
Each excitation contributes to the energy by $|\nu_i-\12|$.
  The energies of the lowest excitations  are
\beq
E=\frac{1}{4}\pm\12\left|\nu_1-\12\right|
   \pm\12 \left|\nu_2-\12\right|
   \pm\12 \left|\nu_3-\12\right| \;\;.
\eeq
In fact, 
these eight states may be classified into two groups each consisting of
 four states.  The  one group has energies
\beq
E^+=\left\{ \ba{l}
\12(-\nu_1-\nu_2-\nu_3+2)\\
\12(\nu_1+\nu_2-\nu_3)\\
\12(\nu_1-\nu_2+\nu_3)\\
\12(-\nu_1+\nu_2+\nu_3)
\ea\right. \;\;\;,
\label{eq:d6energy}
\eeq
and  the other group has energies
\beq
E^-=\left\{ \ba{l}
\12(\nu_1-\nu_2-\nu_3+1)\\
\12(-\nu_1+\nu_2-\nu_3+1)\\
\12(-\nu_1-\nu_2+\nu_3+1)\\
\12(\nu_1+\nu_2+\nu_3-1)
\ea\right.  \;\;.
\eeq
By definition,  the GSO projection keeps
 the states with energies $E^+$ and projects out the states
 with energies $E^-$. 
With this choice, we see if some light states exist.

When $Pf(B)$ is positive, namely, either all of $b_i$ s are positive, or two
 are negative and the one is positive, $E^+$ is about $\pm\12$.
 We have a tachyonic state and three excited states, all 
with mass squared being of order $1$ in the unit of $1/\ap$. 
 These states do not survive  the $\ap\to 0$ limit. 
When $Pf(B)$ is negative, we have two possibilities: either all of $b_i$ s are 
negative, or two of them are positive and the one is negative.

 In the former case,
\beq
\label{eq:d6+}
E^+\approx\left\{ \ba{l}
\12(\frac{1}{\pi b_1}+\frac{1}{\pi b_2}+\frac{1}{\pi b_3}+2)\\
\12(-\frac{1}{\pi b_1}-\frac{1}{\pi b_2}+\frac{1}{\pi b_3})\\
\12(-\frac{1}{\pi b_1}+\frac{1}{\pi b_2}-\frac{1}{\pi b_3})\\
\12(\frac{1}{\pi b_1}-\frac{1}{\pi b_2}-\frac{1}{\pi b_3})
\ea\right. \;\;,
\eeq
 in the zero-slope limit.
The first state in eq. (\ref{eq:d6+}) coming from the three-particle excitation
 is very heavy and the remaining three states coming 
from the one-particle excitations are light. In fact, we have three more light
 states coming from the one-particle excitations which  carry energies
\beq
\12\left(-\frac{3}{\pi b_1}-\frac{1}{\pi b_2}
         {}-\frac{1}{\pi b_3}\right)\;, \hspace{3ex}
\12\left(-\frac{1}{\pi b_1}-\frac{3}{\pi b_2}
         {}-\frac{1}{\pi b_3}\right)\;,\hspace{3ex}
\12\left(-\frac{1}{\pi b_1}-\frac{1}{\pi b_2}
          {}-\frac{3}{\pi b_3}\right)\;\;.
\eeq
Similarly, the lowest modes of the NS fermions from the directions
without $B_{ij}$ carry energy $\12$. These give rise to light states with
 energy
\beq
\label{eq:nsl}
E_i=\12(\nu_1+\nu_2+\nu_3)
  \approx\12 \left(-\frac{1}{\pi b_1}-\frac{1}{\pi b_2}
                   {}-\frac{1}{\pi b_3}\right)\;\;.
\eeq
One can also act on these states with an arbitrary polynomial  consisting of
 the lowest bosonic creation operators $\a_{\nu_i}(i=1,2,3)$. This
gives us a large number of states with finite energy in the $\ap\to 0$ limit
 as each $\a_{\nu_i}$ has energy
$\nu_i\approx -\frac{1}{\pi b_i}$.

 In the latter case, we may take $b_1, b_2$ positive and 
$b_3$ negative   without losing generality.   We obtain
\beq
E^+\approx \left\{ \ba{l}
\12(\frac{1}{\pi b_1}+\frac{1}{\pi b_2}+\frac{1}{\pi b_3})\\
\12(-\frac{1}{\pi b_1}-\frac{1}{\pi b_2}+\frac{1}{\pi b_3}+2)\\
\12(-\frac{1}{\pi b_1}+\frac{1}{\pi b_2}-\frac{1}{\pi b_3})\\
\12(\frac{1}{\pi b_1}-\frac{1}{\pi b_2}-\frac{1}{\pi b_3})
\ea\right. \;\;.
\eeq
Obviously the second state comes from the three-particle excitations and
 is very heavy. The other three states
are light.  Three more light states come from the one-particle excitations with energies
\beq
\12\left(\frac{3}{\pi b_1}+\frac{1}{\pi b_2}-\frac{1}{\pi b_3}\right)\;,
\hspace{3ex}
\12\left(\frac{1}{\pi b_1}+\frac{3}{\pi b_2}-\frac{1}{\pi b_3}\right)\;,
\hspace{3ex}
\12\left(\frac{1}{\pi b_1}+\frac{1}{\pi b_2}-\frac{3}{\pi b_3}\right)\;\;.
\eeq
  As in eq. (\ref{eq:nsl}), we have
\beq
E_i\approx\12
 \left(\frac{1}{\pi b_1}+\frac{1}{\pi b_2}-\frac{1}{\pi b_3}\right)\;\;.
\eeq
The lowest bosonic modes are   $\a_{-1+\nu_1}, \a_{-1+\nu_2}$
and $\a_{\nu_3}$, each of which carries energy 
\beq
\frac{1}{\pi b_1}\;, \hspace{3ex}\frac{1}{\pi b_2}\; ,
 \hspace{3ex} {\rm and}\;  -\frac{1}{\pi b_3}
\eeq
respectively.  These again give rise to a large number of light states.
  We conclude that, in the $Pf(B)<0$ case, there always exist a large number
 of light states in the $\ap \to 0$ limit. 


  The D0-D8 system with $B_{ij}$ field can be studied in the same way.
 In the NS sector, the ground state energy is 
\beq
E_0=\12-\12\sum_{i=1}^{4}\left|\nu_i-\12\right|  \;\;.
\eeq
and the lowest excitations have energies
\beq
E=\12+\12\sum_{i=1}^{4}(\pm)\left|\nu_i-\12\right|  \;\;.
\eeq
These sixteen states including the ground state can be classified into
 two groups each  consisting of eight states. The energies  are 
\beq
E^+=\left\{ \ba{l}
\12(-\nu_1 -\nu_{2}-\nu_{3}-\nu_{4}+3)\\
\12(\nu_1+\nu_2-\nu_3-\nu_4+1)\\
\12(\nu_1-\nu_2+\nu_3-\nu_4+1)\\
\12(\nu_1-\nu_2-\nu_3+\nu_4+1)\\
\12(-\nu_1+\nu_2+\nu_3-\nu_4+1)\\
\12(-\nu_1+\nu_2-\nu_3+\nu_4+1)\\
\12(-\nu_1-\nu_2+\nu_3+\nu_4+1)\\
\12(\nu_1+\nu_2+\nu_3+\nu_4-1)
\ea \right. \;,
\hspace{5ex}
E^-=\left\{ \ba{l}
\12(\nu_1-\nu_2-\nu_3-\nu_4+2)\\
\12(-\nu_1+\nu_2-\nu_3-\nu_4+2)\\
\12(-\nu_1-\nu_2+\nu_3-\nu_4+2)\\
\12(-\nu_1-\nu_2-\nu_3+\nu_4+2)\\
\12(\nu_1+\nu_2+\nu_3-\nu_4)\\
\12(\nu_1+\nu_2-\nu_3+\nu_4)\\
\12(\nu_1-\nu_2+\nu_3+\nu_4)\\
\12(-\nu_1+\nu_2+\nu_3+\nu_4)
\ea \right.\;\;.
  \label{eq:d8energy}
\eeq
The GSO projection selects the states with energies $E^+$. It is not
 difficult to find that, for $Pf(B)>0$,
we have no light state and the mass squared of the tachyonic state is of
 order $1$
 in the unit of $1/\ap$.
When $Pf(B)<0$, we always have a large number of light states
in the $\ap\to 0$ limit. Let us first consider the case
with one positive and three negative $b_i$ s. Taking $b_1>0$ and $b_2,b_3,b_4 <0$,
 one can find eight light states
in the lowest excitations with energies:
\bea
\12 \left(\frac{1}{\pi b_1}+\frac{1}{\pi b_2}-\frac{1}{\pi b_3}
  {}-\frac{1}{\pi b_4} \right)~,
& &\12\left(\frac{1}{\pi b_1}+\frac{1}{\pi b_2}
  {}-\frac{1}{\pi b_3}+\frac{1}{\pi b_4}\right)\;,\nonumber \\
\12\left(\frac{1}{\pi b_1}-\frac{1}{\pi b_2}+\frac{1}{\pi b_3}
  {}-\frac{1}{\pi b_4}\right),
& &\12\left(-\frac{1}{\pi b_1}-\frac{1}{\pi b_2}
  {}-\frac{1}{\pi b_3}-\frac{1}{\pi b_4}\right)\;,\nonumber \\
\12\left(\frac{3}{\pi b_1}-\frac{1}{\pi b_2}-\frac{1}{\pi b_3}
   {}-\frac{1}{\pi b_4}\right),
& &\12\left(\frac{1}{\pi b_1}-\frac{3}{\pi b_2}
   {}-\frac{1}{\pi b_3}-\frac{1}{\pi b_4}\right)\;,\nonumber\\
\12(\frac{1}{\pi b_1}-\frac{1}{\pi b_2}-\frac{3}{\pi b_3}
   {}-\frac{1}{\pi b_4}),& 
&\12\left(\frac{1}{\pi b_1}-\frac{1}{\pi b_2}
   {}-\frac{1}{\pi b_3}-\frac{3}{\pi b_4}\right)  \;\;.
\eea
{}From the bosonic sector, we find the single particle excitations with
 energies 
\beq
1-\nu_1\approx \frac{1}{\pi b_1}\;, \hspace{3ex}-\frac{1}{\pi b_2}\;,
\hspace{3ex}-\frac{1}{\pi b_3}\;, \hspace{3ex}-\frac{1}{\pi b_4}\;\;.
\eeq
 For the case with three positive and one negative $b_{i}$s,  for example,
 $b_1,b_2,b_3>0, b_4<0$,
we have 
\bea
\12\left(\frac{1}{\pi b_1}+\frac{1}{\pi b_2}+\frac{1}{\pi b_3}
+\frac{1}{\pi b_4}\right)\;,
& &\12\left(\frac{1}{\pi b_1}+\frac{1}{\pi b_2}
{}-\frac{1}{\pi b_3}-\frac{1}{\pi b_4}\right) \;,\nonumber \\
\12\left(\frac{1}{\pi b_1}-\frac{1}{\pi b_2}+\frac{1}{\pi b_3}
{}-\frac{1}{\pi b_4}\right)\;,
& &\12(-\frac{1}{\pi b_1}+\frac{1}{\pi b_2}
+\frac{1}{\pi b_3}-\frac{1}{\pi b_4})\;,\nonumber \\
\12\left(\frac{3}{\pi b_1}+\frac{1}{\pi b_2}+\frac{1}{\pi b_3}
{}-\frac{1}{\pi b_4}\right)\;,
& &\12\left(\frac{1}{\pi b_1}+\frac{3}{\pi b_2}
+\frac{1}{\pi b_3}-\frac{1}{\pi b_4}\right)\;,\nonumber\\
\12\left(\frac{1}{\pi b_1}+\frac{1}{\pi b_2}+\frac{3}{\pi b_3}
{}-\frac{1}{\pi b_4}\right)\;,
& &\12\left(\frac{1}{\pi b_1}+\frac{1}{\pi b_2}
+\frac{1}{\pi b_3}-\frac{3}{\pi b_4}\right)\;\;.
\eea
{}From the bosonic sector,
we obtain the one-particle excitations with energies 
\beq
    \frac{1}{\pi b_1}\;, \hspace{3ex}\frac{1}{\pi b_2}\;,
\hspace{3ex}\frac{1}{\pi b_3}\;,\hspace{3ex}-\frac{1}{\pi b_4}\;\;,
\eeq
  which give rise to a large number of light states in the $\ap\to 0$ limit.


\section{The Relation with the Branes at Angles}

It has been realized for some time 
that the system of the two D-branes at angles is T-dual to 
the $p$-$p^\prime$ system
with background ${\cal F}=F+B$ \cite{Berkooz, Leigh}. 
For example,
in the case of D2-D2 at angles, the BPS configuration is T-dual
either to the D0 bound to D4 as a self-dual instanton
or to the $\overline{{\rm D}0}$ bound to D4 as an anti-self-dual
instanton.
Now working on the $p$-$p^\prime$ system in
the presence of $B_{ij}$ field, we reconsider its relation
with the configuration of the branes
at angles, focusing on the  BPS configuration, the existence of light
states and the moduli of noncommutative instantons. The properties of 
the system of branes at angles have also been discussed in \cite{Jabari}. 

In the generic $p$-$p^\prime$ system with $B_{ij}$ field, 
the boundary condition of the $p$-$p^\prime$ open string is
\beq
\p_\tau x^j|_{\s=0}=g_{ij}\p_\s x^j+2\pi i\ap
 B_{ij}\p_\tau x^j|_{\s=\pi}=0 \;\;.
\eeq

Let us first consider the D0-D2 case, {\it i.e.}\/\ $i,j=1,2$,
as the boundary conditions of the generic $p$-$p^{\prime}$
system reduce to those of this case.
 Via T-duality,  we will relate this case to 
 the system of two D1-branes intersecting at an angle. 

We study an open string ending on two D1-branes at
an angle $\phi$ in the $(x^{1},x^{2})$-plane.
We take one of them to be aligned along the $x^{1}$ direction.
In terms of the coordinates $Z=x^1+ix^2$, 
the boundary conditions are \cite{Pol} 
\bea
\label{eq:bc}
\s=0\;, &\p_\s {\rm Re}(Z)=\p_\tau {\rm Im}(Z)=0 \;\;, \nonumber \\
\s=\pi\;, &\p_\s {\rm Re}\left[\exp(i\phi) Z \right]
             =\p_\tau {\rm Im}\left[\exp(i\phi)Z\right]=0\;\;.
\eea
Writing in components, eq. (\ref{eq:bc})  becomes
\bea
\s=0, & \p_\s x^1=0, \p_\tau x^2=0 \;\;, \nonumber \\
\s=\pi, & \p_\s (x^2+b x^1)=0  \;\;, \nonumber \\
 &\p_\tau(bx^2-x^1)=0 \;\;,
\eea
where
\beq
\label{eq:tan}
\tan\phi=-\frac{1}{b} \;\;. 
\eeq
After taking T-duality in the $x^1$ direction 
\beq
\p_\s x^1 \leftrightarrow   -i\p_\tau x^1  \;\;,
\eeq
the boundary condition becomes identical to that of the D0-D2
 system with $B_{ij}$ (\ref{eq:d2bc}), which we wanted to show.

From eq.\ (\ref{eq:tan}), we obtain
\beq
\label{eq:choice}
\phi=\left\{\ba{l}
\nu \pi\;\;,\\
\nu\pi -\pi \;\;.
\ea\right.
\eeq
This means  that we have two choices of angles corresponding to
the same $b$ or $\nu$. 
Which identification we should adopt depends on
the ways   the GSO projection is made.
For the systems of D$p$-D$p$ at angles,
we take the GSO projection so that when all the
angles $\phi_{i}$ vanish we have supersymmetric systems,
{\it i.e.}\/\ the systems of D$p$-D$p$ not D$p$-$\overline{{\rm D}p}$.
We have to fix the identfication between the angles $\phi_{i}$
and $\nu_{i}$ in such a way as the situation is cosistent
with the GSO projections taken in the last section.

In what follows we will fix the relations between
the angles and $\nu$'s for all cases discussed in the last
section.
We will show that the energies of the ground states surviving
the GSO projection that we obtained in the last section
are described in the forms related to the BPS conditions
for the systems of the D-branes at angles.


Let us look at the D0-D2 system more closely.
Taking T-duality, we obtain D1-D1 at an angle $\phi$.
Since the system must be supersymmetric when $\phi = 0$,
the GSO projection (\ref{eq:pm}) tells us to
identify $\phi$ with $\nu$ in the following way,
\beq
\phi=\nu\pi.
\label{eq:phinu-1}
\eeq
It follows that the energy $E^{-}$ in (\ref{eq:pm})
of the sate surviving the GSO projection is expressed as
\begin{equation}
 E^{-} = -\frac{1}{2\pi}\phi~.
\end{equation}
For $b<0$ we have $\phi \approx 0$ in the zero-slope limit.
The system becomes almost sypersymmetric, 
{\it i.e.}\/\ the D1-D1 system, and the ground state is nearly
massless. The bosonic fluctuations generated by $\alpha_{\nu}$'s
with energy $\nu = \phi/\pi \approx 0$ give us a large number of light
sates in this limit.
On the other hand, for $b>0$ we have $\phi \approx \pi$.
The system becomes non-supersymmetric, {\it i.e.}\/\ almost
D1-$\overline{\rm D1}$ system, and the ground state becomes tachyonic.

As we pointed out in the last section,
when $\nu$ varies from $0$ to $1$
the level crossing happens between the ground state
and the first excited state. This yields the  apparent
flip of the GSO projection. Combining this fact and
the identification (\ref{eq:phinu-1}), we are able to explain how
we can convert the D1-D1 system ($\phi =0$) continuously
to the D1-$\overline{\rm D1}$ system ($\phi=\pi$)
while they have the GSO projections opposite to each other.



As the next example, let us consider the D0-D4 case with $B$ field.
When we take T-duality in the $x^{1}$- and the $x^{3}$-directions,
the system becomes D2-D2 at angles $\phi_{1}$ 
and $\phi_{2}$ in $(x^{1},x^{2})$- and $(x^{3},x^{4})$-planes
respectively.
For the consistency with the GSO projection
(\ref{eq:d2energy})
we should identify the angles with $\nu$'s as follows:
\begin{equation}
 \phi_{1}=\nu_{1}\pi~, \qquad \phi_{2}=\nu_{2}\pi - \pi~\label{D0D4}.
\end{equation}
Among the supersymmetry conditions for the system of D2-D2 at
angles
\begin{equation}
 \phi_{1}+\phi_{2}\equiv 0~,\quad
 \phi_{1}-\phi_{2}\equiv 0 \quad (\mbox{mod $2\pi$})~,
\label{eq:bpsd2d2}
\end{equation}
the former equation is that for the present D0-D4 system.
The energies $E^{+}_{\pm}$ in (\ref{eq:d2energy})
of the states surviving the GSO projection are expressed as
\beq
E^+_{\pm} = \pm \frac{1}{2\pi}\left( \phi_1+\phi_2 \right)~.
\eeq
For $Pf(B)>0$
we have $\phi_1+\phi_2 \approx \pm\pi$ in the zero-slope limit.
We have an almost D2-$\overline{\rm D2}$ system and
the ground sate is tachyonic.
For $Pf(B)<0$  we obtain $\phi_1+\phi_2\approx 0$ in the
zero-slope limit.
We obtain an almost D1-D1 system.
The lowest energy states are almost massless and
the bosonic fluctuations with energies proportional
to the very small angles give us many light states in this limit. 

On the other hand, if we consider the $\overline{\rm D0}$-D4 system,
we should choose the GSO projected states with energies 
$E^-_{\pm}=\pm\12(\nu_1-\nu_2)$,
which is proportional to $\phi_1-\phi_2$ under identification
$\phi_i=\nu_i\pi$. 
Note that the condition for supersymmetry now turns out to 
be the latter equation in eq.\ (\ref{eq:bpsd2d2}).
Therefore, only when $Pf(B)>0$, we have a large number of light states
coming from the $\overline{\rm D0}$-$\overline{\rm D0}$ pairs.
It can be interpreted as the fluctuations around the supersymmetric 
configuration in the picture of branes at angles.

Another point which is worth mentioning is that the supersymmetric
condition,
either of the form of $\phi_1+\phi_2=0$ for D0-D4 or
of the form of $\phi_1-\phi_2=0$ for $\overline{\rm D0}$-D4,
is closely related to the moduli of instantons or anti-instantons on 
noncommutative $R^4$. 
As argued in \cite{SW}, in the case of D0-D4 with $Pf(B)<0$,
the tachyon mass
squared vanishes when $b_1+b_2=0$ and the system is supersymmetric.
From the dual picture of branes at angles,
this correspond to $\phi_1+\phi_2=0$ under our identification
(\ref{D0D4}). 
As for $\overline{\rm D0}$-D4, when $b_1=b_2$,
the system becomes supersymmetric and BPS,
representing a point on noncommutative 
anti-instanton moduli space.

Next we consider the D0-D6 system with $B$ field.
When we take T-duality in three of the spatial directions
along the D6-brane worldvolume,
we obtain the system of D3-D3 at angles $\phi_{1}$,
$\phi_{2}$ and $\phi_{3}$.
By repeating the considerations on supersymmetry and
the GSO projection similar to those of the D0-D2 and the D0-D4 cases,
we find that we should choose one of the following identifications:
\begin{eqnarray}
&& \phi_{i}=\nu_{i}\pi \quad (i=1,2,3)~,\nonumber\\
&&  \phi_{1}=\nu_{1}\pi~, \quad \phi_{2}=\nu_{2}\pi-\pi~,
\quad \phi_{3}=\nu_{3}\pi-\pi~,
\end{eqnarray}
and its permutations.
In both identifications,
the energies $E^{+}$ of the states surviving the GSO projection,
listed in (\ref{eq:d6energy}), are expressed as
\begin{equation}
 2 \pi E^{+} \equiv 
  \left\{
    \begin{array}{l}
     (-\phi_{1}-\phi_{2}-\phi_{3})\\
     (\phi_{1}+\phi_{2}-\phi_{3})\\
     (\phi_{1}-\phi_{2}+\phi_{3})\\
     (-\phi_{1}+\phi_{2}+\phi_{3})
  \end{array}
\quad \mbox{(mod $2\pi$)}~.
\right.
\end{equation}
{}From this we find that when the states become massless,
the condition that supersymmetry for the system of D3-D3 at
angles be unbroken is satisfied.
This means that at least one of
the following equations holds:
\begin{equation}
 \phi_{1} \pm \phi_{2} \pm \phi_{3} \equiv 0
 \quad \mbox{(mod $2\pi$)}~.
\end{equation}


Finally let us consider the case of D0-D8.
Taking T-duality in four of the spatial directions
along the D8-brane worldvolume,
we obtain the system of the D4-D4 at angles $\phi_{i}$
$(i=1,\ldots,4)$.
In this case we should choose one of the following identifications:
\begin{eqnarray}
&& \phi_{1}=\nu_{1}\pi-\pi~, \qquad
   \phi_{i}=\nu_{i}\pi \quad (i=2,3,4)~,\nonumber\\
&&  \phi_{1}=\nu_{1}\pi~, \qquad
  \phi_{i}=\nu_{i}\pi-\pi \quad (i=2,3,4)~,
\end{eqnarray}
and the permutations of them.
As in the three previous cases,
supersymmetry is partially recovered
when some states of energy $E^{+}$ listed in eq.\ (\ref{eq:d8energy})
become massless. Therefore
at least one of the following conditions for supersymmetry 
is satisfied:
\bea
 \phi_{1}+\phi_{2}+\phi_{3}+\phi_{4}\equiv 0, 
\quad (\mbox{mod $2\pi$})~,& &
\phi_{1}+\phi_{2}-\phi_{3}-\phi_{4}\equiv 0, 
\quad (\mbox{mod $2\pi$})~,\nonumber\\
\phi_{1}-\phi_{2}-\phi_{3}+\phi_{4}\equiv 0, 
\quad (\mbox{mod $2\pi$})~,& &
\phi_{1}-\phi_{2}+\phi_{3}-\phi_{4}\equiv 0, 
\quad (\mbox{mod $2\pi$})~.
\eea
Obviously, for $Pf(B)<0$ a large number of light states
appear in the zero-slope limit
and for $Pf(B)>0$ the system is not supersymmetric and the
T-dualized picture is almost D4-$\overline{\rm D4}$ pair.

\section{One-Loop Amplitudes and the Branes with Relative Motion}

In this section we will relate the $p$-$p^\prime$
system in the presence of $B_{ij}$ field with moving D-branes. 
As is well known, the system of the D-branes with relative motion
has a connection 
with D-branes in a constant electric background \cite{Bachas1, Bachas2,Jabari}.
In the last section we considered the D-branes in a constant magnetic
field, which can be related to D-branes at angles by T-duality.
Therefore in a generic $B_{\mu\nu}$ background, the $p$-$p^{\prime}$ system
can be either thought of as D-branes with relative motion or
the ones with relative orientation.
This fact reminds us of the motion of a charged particle
in the presence of constant electric field or magnetic field. 

Suppose that the two D-branes move with a relative velocity $v$.
The mode expansion of an open string satisfying the appropriate boundary
condition has almost the same form as eq.\ (\ref{eq:mode})
but with a pure imaginary parameter:
\beq
i\epsilon \equiv \frac{\arctan(v)}{\pi}=\nu~,  \label{eq:velocity}
\eeq
where $\nu$ is the parameter in the mode expansion (\ref{eq:mode}).
This identification shows that the two problem can
be treated in one way. 


As an illustration, let us consider the open string one-loop
vacuum amplitude for the D0-D2 case. 
It takes the form
\bea
A&\sim&\int^\infty_0\frac{d^{(\sharp NN+1)}k}{(2\pi)^{(\sharp NN+1)}}
       \sum_i\int^\infty_0\frac{dt}{t}
         \exp\left[-2\pi\ap t \left(k^2+M^2_i \right) \right] \nonumber\\
&=&\sum_i\int^\infty_0\frac{dt}{t}(32\pi^2\ap t)^{-\frac{1}{2}}
   \exp\left( -2\pi\ap tM^2_i \right)  \;\;,
\eea
where
\beq
M^2_i=\frac{y^2}{4\pi^2\alpha^{\prime 2}}
   +\frac{1}{\ap}\sum(  {\rm oscillators}) \;\;,
\eeq
with $y$ being the distance between two D-branes.
We denote by $\sharp NN$
the number of directions in which both ends of the open string obey
the Neumann boundary condition. Here $\sharp NN=0$.
Then the amplitude is 
\bea
A \sim \int^\infty_0\frac{dt}{t}
\left(32\pi^2\ap t\right)^{-\frac{1}{2}}
 \exp\left(-\frac{ty^2}{2\pi\ap}\right) \cdot
B\times F \;\;,
\label{eq:amp}
\eea
with
\bea
B&=&\left(\eta(it)\right)^{-6} 
  \left\{
  {}-i\,e^{\nu^2\pi t}\,\frac{\eta(it)}{\vt_{11}(i\nu t;it)}
   \right\}  \;\;, \\
F&=&\12 \left\{\prod^4_{a=1}{Z^0}_{0}(\nu_a;it)-
\prod^4_{a=1}{Z^0}_{1}(\nu_a;it)-\prod^4_{a=1}{Z^1}_{0}(\nu_a;it)
{}-\prod^4_{a=1}{Z^1}_{1}(\nu_a;it)\right\}\;\;,  \\
&=&\left[Z^1_{~1}\left(\frac{\nu}{2};it\right)\right]^4 \;\;,
\eea
where $\nu=\nu_1\neq 0$, and $\nu_{2,3,4}=0$.
 $B$ is the contribution from the bosonic oscillators
 and $F$ is the one from fermionic parts.  We have introduced
\beq
{Z^\a}_{\b}(\nu; it)=
\frac{\vt_{\a\b}(i\nu t; it)}{e^{\nu^2\pi t}\eta(it)} \;\;.
\eeq

Let us focus on the long range potential
between the branes.
We approximate eq. (\ref{eq:amp})
in the limit of $\ap\to 0$ and small $t$. Using the modular 
property of Jacobi function, we find that
\bea
\vt_{11}(i\nu t; it)&=&-it^{-\12}\exp(\nu^2\pi t)
    \vt_{11}\left(\nu;\frac{i}{t}\right)~, \\
\vt_{11}\left(\nu;\frac{i}{t}\right)
   &=&-2q^{\frac{1}{8}}\sin(\nu\pi)\prod_m(1-q^m)(1-zq^m)(1-z^{-1}q^m)~,
\eea
where
\beq
q=e^{-\frac{2\pi}{t}}~, \qquad z=e^{2i\pi\nu}~.
\eeq
In the small $t$ limit, this leads to
\beq
\vt_{11}(i\nu t; it)\approx 2it^{-\12}
 \exp(\nu^2\pi t)\sin(\nu\pi)\exp\left(-\frac{\pi}{4t}\right)~.
\eeq
Similary, by using the modular transformation of the Dedekind
eta finction, we find that in this limit
\begin{equation}
 \eta (it) \approx t^{-\frac{1}{2}}
      \exp\left(\frac{\pi}{12 t}\right)~.
\end{equation}
Gathering all of the contributions from the bosonic and
the fermionic sectors, we obtain
\beq
A \propto \frac{\sin^{4}\left(\frac{\nu\pi}{2}\right)}{\sin\nu\pi}
  \int^{\infty}_{0} dt \;
    \frac{2^{\frac{1}{2}}}{\pi \alpha^{\prime \frac{1}{2}}}\,
     t^{\frac{3}{2}}\,
    \exp\left(\frac{t y^{2}}{2\pi \alpha^{\prime}}\right)~.
\label{eq:amp-2}
\eeq

We can read off the potential in the zero-slope limit
from the amplitude (\ref{eq:amp-2}).
For $b<0$ {\it i.e.}\/\ $\nu\approx 0$, we have a quite small
potential showing that the system is in an
almost supersymmetric configuration.
For $b>0$, $\nu$ tends to $1$.
The potential becomes very large. This indicates that
the system is far from being supersymmetric.

\section*{Acknowlegements:}

We are grateful to T. Yokono for useful discussions.



\begin{thebibliography}{99}

\bibitem{Witten2}E. Witten, "Bound States of Strings and p-Branes",
	hep-th/9510135.
\bibitem{CDS}A. Connes, M.R. Douglas and A. Schwarz, "Noncommutative
	Geometry and Matrix Theory: Compactification 
        On Tori", JHEP {\bf 9802:003}(1998), hep-th/9711162.
\bibitem{DH}M.R. Douglas and C. Hull, "D-branes And The Noncommutative
	Torus", JHEP {\bf 9802:008}(1998), hep-th/9711165.
\bibitem{AS}A. Schwarz, "Morita Equivalence And Duality",
	Nucl. Phys. {\bf B534}(1998)720, hep-th/9805034; 
        B. Pioline and A. Schwarz, "Morita Equivalence and T-duality
       (or $B$ versus $\Theta$)", hep-th/9908019. 
\bibitem{SW}N. Seiberg and E. Witten, "String Theory and Noncommutative
	Geometry", hep-th/9908142. 
\bibitem{Naka}H. Nakajima, "Resolutions of Moduli Spaces of Ideal
	Instantons on $R^4$", in "Topology, Geometry
        and Field Theory"(World Scientific, 1994).
\bibitem{Nek}N. Nekrasov and A. Schwarz, "Instantons On Noncommutative
	$R^4$ And (2,0) Superconformal Field Theory", 
        Comm. Math. Phys. {\bf 198}689, hep-th/9802068.
\bibitem{Witten1}E. Witten, "Small Instanton in String Theory",
	Nucl. Phys. {\bf B460}541, hep-th/9511030.
\bibitem{Douglas}M. Douglas, "Branes Within Branes", hep-th/9512077.



\bibitem{Berkooz}M. Berkooz, M.R. Douglas and R.G. Leigh, "Branes
	Intersecting at Angles", Nucl. Phys. {\bf B480}(1996)265,
	hep-th/9606139.
\bibitem{Leigh}V. Balasubramanian and R.G. Leigh, "D-Branes, Moduli and
	Supersymmetry", Phys. Rev. {\bf D55}(1997)6415, hep-th/9611165.
\bibitem{Jabari}H. Arfaei and M.M. Sheikh-Jabbari, "Different D-brane Interactions", Phys. Lett. {\bf B394}(1997)288, hep-th/9608167.\\
	M.M. Sheikh-Jabbari, "Classification of Different Branes at Angles",
	Phys. Lett. {\bf B420}(1998)279, hep-th/9710121.\\
	M.M. Sheikh-Jabbari, "More on Mixed Boundary Conditions and D-branes Bound States", Phys. Lett. {\bf B425}(1998)28, hep-th/9712199.\\
	D. O'Driscoll, "Towards a Supersymmetic Classification of D-Brane Configurations with Odd Spin Structure", hep-th/9908071.
\bibitem{Pol}J. Polchinski, "String Theory" Vol. II, Cambridge
	University Press(1998).
\bibitem{Callan1}C.G. Callan, C. Lovelace, C.R. Nappi, S.A. Yost,
	"String Loop Corrections To Beta Functions",
        Nucl. Phys. {\bf B288}(1987)525; A. Abouelsaood, C.G. Callan,
	C.R. Nappi and S.A. Yost, "Open Strings In Background Gauge
	Fields", Nucl. Phys. {\bf B280}(1987)599.\\
        E.S. Fradkin and A.A. Tseytlin, "Nonlinear Electrodynamics From
	Quantized Strings", Phys. Lett. {\bf 163B}(1985)123.
\bibitem{Bachas1}C. Bachas, "D-Brane Dynamics", 
        Phys. Lett. {\bf B374}(1996)37, hep-th/9511043.
\bibitem{Lif}G. Lifschytz, "Comparing D-branes to Black-branes",
	Phys. Lett. {\bf B388}(1996)720, hep-th/9604156.
\bibitem{Bachas2}C. Bachas and M. Porrati, "Pair Creation of Open
	Strings in an Electric Field", Phys. Lett. {\bf 296}(1992)77,
        hep-th/9209032.


\end{thebibliography}
\end{document}